\documentclass[11pt,tightenlines,eqsecnum,floats,aps,amsmath,amssymb,nofootinbib,prd,shownopacs,floatfix]{revtex4}
\usepackage{graphicx}
\usepackage{epstopdf}
\usepackage{latexsym}
\usepackage{amssymb}
\usepackage{amsmath}
\usepackage{color}
\usepackage{mathrsfs}
\usepackage{xparse}
\usepackage{float}
\usepackage{mathtools}
\usepackage[center]{subfigure}
\DeclareMathOperator{\sinc}{sinc}

\begin{document}
\renewcommand\arraystretch{2}
\newcommand{\bq}{\begin{equation}}
\newcommand{\eq}{\end{equation}}
\newcommand{\bqn}{\begin{eqnarray}}
\newcommand{\eqn}{\end{eqnarray}}
\newcommand{\nb}{\nonumber}
\newcommand{\lb}{\label}
\newcommand{\cb}{\color{blue}}
\newcommand{\cc}{\color{cyan}}
\newcommand{\cm}{\color{magenta}}
\newcommand{\rc}{\rho^{\scriptscriptstyle{\mathrm{I}}}_c}
\newcommand{\rd}{\rho^{\scriptscriptstyle{\mathrm{II}}}_c} 
\NewDocumentCommand{\evalat}{sO{\big}mm}{%
  \IfBooleanTF{#1}
   {\mleft. #3 \mright|_{#4}}
   {#3#2|_{#4}}%
}
\newcommand{\PRL}{Phys. Rev. Lett.}
\newcommand{\PL}{Phys. Lett.}
\newcommand{\PR}{Phys. Rev.}
\newcommand{\CQG}{Class. Quantum Grav.}
\newcommand{\parallelsum}{\mathbin{\!/\mkern-5mu/\!}}

\title{Loop quantum cosmology and its gauge-covariant avatar: \\ a weak curvature relationship}

\author{Bao-Fei Li$^{1,2}$}
\email{baofeili1@lsu.edu}
\author{Parampreet Singh$^1$}
\email{psingh@lsu.edu}
\affiliation{
$^{1}$ Department of Physics and Astronomy, Louisiana State University, Baton Rouge, LA 70803, USA\\
$^{2}$ Institute for Theoretical Physics $\&$ Cosmology, Zhejiang University of Technology, Hangzhou, 310032, China\\}

\begin{abstract}

We explore the relationship between the effective dynamics in standard loop quantum cosmology (LQC) based on holonomies and triads obtained from gauge-fixing fluxes, 
and a  modification of LQC  based on holonomies and gauge-covariant fluxes (referred to as gLQC).  Both the models yield singularity resolution via a bounce because of non-perturbative quantum geometric effects resulting in a maximum for energy density. In LQC, the bounce is extremely well captured by a $\rho^2$ term in energy density with a negative sign which emerges as a non-perturbative modification to the classical Friedmann and Raychaudhuri equations. But, details of such modifications in gLQC have remained hidden due to  an arduous nature of gauge-covariant flux modifications which  do not allow writing above equations in a closed form. To extract these modifications we explore the large volume,  weak curvature limit for matter with a fixed equation of state and obtain higher order corrections to the classical theory. We find that in the  weak curvature limit of gLQC, in the post-bounce branch, the first order correction beyond classical theory fully recovers the form of  modified Friedmann and Raychaudhuri equations of LQC. In contrast, due to an asymmetric bounce in gLQC, the weak curvature limit of the pre-bounce branch exhibits a novel structure with a $\rho^{3/2}$ term as a first order correction beyond classical theory while the $\rho^2$ term appears as a  second order correction. Our work shows that gLQC has a far richer structure which includes the form of dynamical equations with non-perturbative modifications in LQC  in its weak curvature limit. This indicates that more general loop quantizations of cosmological sectors can reveal LQC at some truncation, and possibly there exist a  tower of potentially interesting higher order modifications from quantum geometry which are hidden in the setting of LQC.
\end{abstract}

\maketitle

\section{Introduction}
\label{introduction}
\renewcommand{\theequation}{1.\arabic{equation}}\setcounter{equation}{0}

An important issue in understanding quantum geometric effects originating from loop quantum gravity (LQG) in the very early universe is the robustness of physical implications derived from the quantization of cosmological models which implement its techniques in one way or another. Loop quantum cosmology (LQC) \cite{Ashtekar:2011ni} is a rigorous quantization of symmetry-reduced cosmological models based on LQG, which has been used to explore in detail consequences for resolution of singularities \cite{Singh:2009mz,Singh:2014fsy} and various phenomenological consequences \cite{Agullo:2016tjh}. A key prediction of the model is big bounce which occurs when energy density of the universe reaches a Planckian value \cite{Ashtekar:2006rx,Ashtekar:2006uz,Ashtekar:2006wn,Ashtekar:2007em}. It turns out that underlying quantum dynamics in LQC can be captured extremely well by using an effective spacetime description \cite{Diener:2014mia, Diener:2017lde,Singh:2018rwa} which results in a  modified set of Friedmann and Raychaudhuri equations encoding quantum geometry effects through a $\rho^2$ modification in energy density. In the derivation of the effective Hamiltonian from quantum theory, this $\rho^2$ modification, with a negative sign, captures the entire non-perturbative modification for sharply peaked states and thus explains entire dynamics including Planck regime. Even for highly quantum states, where quantum fluctuations start playing a significant role and lower the bounce density the $\rho^2$ modification describes the kinetic dominated bounce extremely well \cite{Corichi:2011sd,Diener:2014hba, Ashtekar:2015iza}. In this sense, at an effective level, the physics of Planck regime extracted from LQC, to some extent is the physics of a $\rho^2$ modified quantum cosmological model.

In spite of the success of LQC in the last two decades, open questions remain on whether it reliably captures the cosmological sector of LQG \cite{Engle:2007qh,Brunnemann:2007du}. There are various quantization ambiguities in the construction of the theory and recovering its cosmological sector and it is important to understand the physical consequences of the regularization prescriptions and refinements.  
Attempts to address this issue come from both a top-down as well as a bottom-up approaches. The top-down approaches attempt to identify a suitable cosmological sector from LQG, such as using quantum-reduced loop gravity \cite{Alesci:2013xd}, coherent states \cite{Dapor:2017rwv} or the path integral approach \cite{Han:2019feb}. On the other hand, the bottom-up approaches aim to generalize LQC with implementing different regularizations of the Hamiltonian constraint \cite{Yang:2009fp, Li:2018opr, Assanioussi:2018hee} and implementing novel refinements used in LQG (see for eg \cite{Liegener:2019dzj, Liegener:2019ymd, Liegener:2019zgw}). In addition to these, attempts have been made to extract modified Friedmann equation of LQC using spinfoam cosmology \cite{Bianchi:2010zs} and group field theory techniques \cite{Gielen:2013naa}. From a phenomenological viewpoint, above issue can be expressed as whether these generalizations or refinements result in a modified Friedmann dynamics as in LQC and how similar is the physics of the Planck regime. Recently, this issue has been investigated in some detail for modified versions of LQC using Thiemann regularization of the Hamiltonian constraint and detailed phenomenological investigations for inflationary models and CMB have been made \cite{Agullo:2018wbf,Li:2018fco,Li:2019ipm,Li:2019qzr,Li:2020mfi,Li:2021mop}. The modified Friedmann and Raychaudhuri equations in these models are distinct from LQC since they contain higher order terms than $\rho^2$ in energy density, but the qualitative results  for the post-bounce branch turn out to be similar for inflationary models. As a result, in these modified versions of LQC, scale-invariant primordial scalar and tensor power spectrum can be obtained as well for the short wavelength modes \cite{Li:2019qzr,Li:2020mfi}. The potentially observable signals which can be used to differentiate these models from LQC are expected to come from the long wavelength modes which are outside the Hubble horizon. On the other hand, for models with negative potentials such as in Ekpyrotic/cyclic cosmologies, it turns out that an important variant of LQC does not lead to a cyclic universe \cite{Li:2021fmu}.

However, uncovering quantum geometry effects in loop cosmology, especially via a modified Friedmann dynamics can be sometimes arduos.  This has been made evident recently in a variant of LQC where instead of using holonomies of Ashtekar-Barbero connection and gauge-fixed triads, one uses holonomies of connection and ``gauge-covariant" fluxes \cite{Liegener:2019zgw,Liegener:2019dzj}. The motivation for using these fluxes arises from LQG where one does not quantize connection and triad variables directly, but the holonomies of connection along an edge and the fluxes of the triad using a two dimensional surface. It turns out that requiring a closure of Poisson bracket of fluxes violates the Jacobi identity and passage to quantum theory becomes difficult. A way out of this problem is to use a Lie algebra of holonomies and vector fields based on fluxes and then quantize \cite{Ashtekar:1998ak}. However, applying this strategy for recovering a cosmological sector from LQG using suitable coherent states is quite problematic. Instead, it is useful to consider another solution to above problem  which is to use ``gauge-covariant" fluxes using triads and connections \cite{Thiemann:2000bv}. In the symmetry reduced setting using gauge-covariant fluxes implies replacing symmetry-reduced triads $p$ to $p\rightarrow p \sinc^2\left(c\epsilon/2\right)$ \cite{Liegener:2019zgw, Liegener:2019ymd} where the edge length $\epsilon$ is fixed by the quantum geometry. This version of LQC where one uses gauge-covariant fluxes will be referred to as ``gLQC'' to differentiate from LQC. It is distinct from other modified versions of LQC \cite{Li:2018opr} in the sense that it starts with the same regularization of the Hamiltonian constraint. And unlike these variants, the convoluted nature of gauge-covariant fluxes do not allow writing the modified Friedmann equation in a closed form. Although the physical implications of the theory can still be extracted from the Hamilton's equations by numerical methods, the modified Friedmann and Raychaudhuri equations can give a more transparent picture on how some observables (such as the Hubble rate) evolve with other observables (such as the energy density and the pressure), and yield important insights on the nature of quantum gravitational modifications. Due to absence of these equations and intricate nature of quantum geometry effects, physics of gLQC has so far not been explored in detail. It is however known to yield a non-singular bounce with a maximum energy density. The bounce is asymmetric in nature and in the asymptotic large volume limit, classical Friedmann and Raychaudhuri equations are recovered \cite{Liegener:2019ymd}. 

As we will demonstrate in this manuscript, obtaining modified Friedmann and Raychaudhuri equations in presence of gauge-covariant fluxes is quite difficult because  one can not analytically solve for the momentum $b$ (which is conjugate to volume and proportional to the Hubble rate in classical regime) in terms of the energy density from the Hamiltonian constraint. The exception to this only happens when the matter content consists of a positive cosmological constant which results in a constant momentum $b$ and results in the Friedmann and Raychaudhuri equations exactly in the same form of the classical theory albeit with a rescaled cosmological constant.\footnote{The situation in LQC for the positive cosmological constant case is similar \cite{Singh:2009mz}.} 
To understand quantum gravity modifications resulting from gauge-covariant fluxes to Friedmann dynamics for various types of matter content, we examine in detail the weak curvature limit of gLQC and extract higher order modifications to the classical theory coming from quantum geometry. The goal of this manuscript is to understand these particular corrections working in the regime of large volume for matter with a fixed equation of state satisfying weak energy condition which means the limit of weak spacetime curvature where the energy density is much below the Planck scale. It turns out that in gLQC, the energy conservation law holds in terms of the gauge covariant energy density and pressure which are defined in terms of the gauge covariant volume, then it is natural to investigate the asymptotic behavior of the modified Friedmann  and Raychaudhuri equations in terms of the gauge covariant counterparts of the conventional quantities. In practice, we find the asymptotic expansion of the momentum $b$ in terms of the inverse powers of the gauge covariant volume which is required to be consistent with the Hamiltonian constraint and then plug the asymptotic expansion into the Hamilton's equation of the gauge covariant volume. This leads to a final expression of the modified Friedmann and Raychaudhuri equations expressed as a sum of powers in the gauge covariant energy density. Our investigations reveal some surprising results. We find that for the post-bounce branch, in the asymptotic regime of large volume, the first order correction beyond GR in gLQC results in the same form of Friedmann and Raychaudhuri equations for gauge-covariant quantities as in standard LQC which in the latter govern the entire evolution (including bounce). That is, in terms of gauge-covariant variables we recover the  $\rho^2$ term with the same maximum energy density of LQC. This is equivalent to say that an observer living in a universe  governed by gLQC will regard LQC as the first order quantum correction to general relativity (GR). In this sense, ``LQC Friedmann dynamics" is recovered as a first order correction to GR in gLQC! On the other hand, when the energy density tends to the Planck energy density, the dynamics of gLQC can not be completely captured by the first order correction as the higher-order terms also become important. Therefore, taking into account the contributions of the higher-order terms, the maximum energy density at the bounce in gLQC turns out to be larger than the one in LQC. Finally, in the pre-bounce branch, given the asymmetry of bounce, the $\rho^2$ term is obtained but at the second order beyond GR while an additional $\rho^{3/2}$ term shows up as the next-to-leading-order correction beyond GR. 

This manuscript is organized as follows. In Sec. \ref{sec:review}, we briefly review the effective dynamics of gLQC and observe that unlike  LQC, the modified Friedmann and Raychaudhuri equations can not be expressed in closed forms in gLQC. We also present a special case of positive cosmological constant which allows a closed form of modified Friedmann equation in presence of gauge-covariant fluxes. In Sec. \ref{sec:asymptotic gauge covariant LQC}, we study the asymptotic behavior of the modified Friedmann and Raychaudhuri equations in the expanding and the contracting phases of gLQC when the universe is filled with a perfect fluid with a fixed equation of state. Although no closed forms of these equations are available, one can still use the large volume expansion of the momentum $b$ to express these equations by a sum of powers of the energy density. The asymptotic forms of the modified Friedmann and Raychaudhuri equations in terms of the gauge covariant quantities as well as the conventional quantities are presented and discussed. In Sec. \ref{sec:summary}, we summarize our main results. 
In the following, we use the Planck units $\hbar=c=1$ and keep Newton's constant $G$ implicit in the constant $\kappa$ with $\kappa=16 \pi G$.

\section{The effective dynamics of \MakeLowercase{g}LQC: Preliminaries}
\label{sec:review}
\renewcommand{\theequation}{2.\arabic{equation}}\setcounter{equation}{0}

We start by discussing some important features of the effective dynamics of gLQC  in a spatially-flat FLRW spacetime. As always assumed in LQC, we assume that the effective Hamiltonian of gLQC can be derived from the underlying quantum theory with a suitable choice of semi-classical states. Given the homogeneity and isotropy of the background spacetime, the classical phase space consists of the symmetry reduced connections and triads $c$ and $p$ -- the same variables before quantization in LQC. But in contrast to the latter, at the level of the quantum and the effective dynamics, quantum geometry effects in gLQC are not only encoded through holonomy of connections but also gauge-covariant fluxes. These effects can be understood via a ``polymerization" of the connection and triads. Since the connection is polymerized in the same way as in LQC, one can obtain the effective Hamiltonian in gLQC  using in addition $p\rightarrow p \sinc^2\left(c\epsilon/2\right)$ \cite{Liegener:2019zgw, Liegener:2019ymd}. Here  $\epsilon$ is the edge length determined following the improved dynamics (or the $\bar \mu$ scheme) in  LQC \cite{Ashtekar:2006wn}, and is given by the physical area of the loop. In particular,  $\epsilon=\sqrt{\Delta/p}$ where $\Delta$ denotes the minimal eigenvalue of the area operator in LQG. 

Since $\epsilon$ directly depends on the triad variable $p$, as in LQC, it is more convenient to employ an equivalent set of canonical variables which are defined via $v=|p|^{3/2}$ and $b=c|p|^{-1/2}$ \cite{Ashtekar:2007em},  
with their fundamental Poisson bracket given by $\{b, v\}=4\pi G\beta$. Here $\beta$ is the Barbero-Immirzi parameter whose value can be fixed using black hole thermodynamics in LQG. For numerical purposes, we choose $\beta = 0.2375$, as in 
 the previous works in LQC. For the matter sector,  we consider a perfect fluid with a fixed equation of state $w$, and energy density $\rho = \rho_0 (v/v_0)^{-(1 + w)}$ with $\rho_0$ and $v_0$ as constants fixed by initial conditions. In the following we take $v_0 = 1$.  
 
 The effective Hamiltonian constraint takes the form \cite{Liegener:2019ymd,Liegener:2019dzj}
\bq
\lb{gauge covariant ham}
\mathcal H_{\mathrm{g.c.}} =-\frac{6v}{\kappa \beta^2\lambda^2}\sin^2\left(\lambda b\right)\sinc\left(\frac{\lambda b}{2}\right) +\mathcal H^{\mathrm{g.c.}}_m, 
\eq
where $\lambda=\sqrt{\Delta}$ and the matter part of the Hamiltonian constraint is given by 
\bq
\mathcal H^{\mathrm{g.c.}}_m=\rho_0 v^{-w}\sinc^{-3w}\left(\frac{\lambda b}{2}\right),
\eq
which describes a perfect fluid with a fixed equation of state $w$. This is obtained noting that at the classical level the matter part of the Hamiltonian constraint is ${\cal H}_m = \rho v$, and using  $v\rightarrow v \sinc^3\left(\lambda b/2\right)$. 
Above, the index  `g.c.'  denotes gauge-covariant quantities which, as compared with their counterparts in LQC, acquire additional contributions characterized by the $\sinc$ terms arising from the gauge-covariant fluxes. An important 
variable of interest is gauge-covariant volume which is related with the conventional volume via $v_{\mathrm{g.c.}}=v\sinc^3\left(\lambda b/2\right)$.  Correspondingly, all the physical quantities that are functions of the volume or its time derivatives have gauge-covariant analogues.  For instance, we can define the gauge-covariant Hubble rate as 
\bq
\lb{hubble rate}
H_{\mathrm{g.c.}}=\frac{\dot v_{\mathrm{g.c.}}}{3v_{\mathrm{g.c.}}},
\eq
which is related with the conventional Hubble rate via
\bq
\lb{conversion between h}
H_\mathrm{g.c.}=H+\frac{\mathrm d\sinc\left(\lambda b/2\right)/\mathrm d t}{\sinc\left(\lambda b/2\right)}.
\eq
Similarly, the gauge-covariant energy density and pressure are given by
\bqn
\lb{density}
\rho_{\mathrm{g.c.}}&\equiv& \frac{\mathcal H^\mathrm{g.c.}_m}{v_{\mathrm{g.c.}}}=\frac{\rho_0}{v^{1+w}_{\mathrm{g.c.}}},\\
\lb{pressure}
P_{\mathrm{g.c.}}&\equiv& -\frac{\partial \mathcal H^\mathrm{\mathrm{g.c.}}_m}{\partial v_{\mathrm{g.c}}}=w\rho_{\mathrm{g.c.}},
\eqn
which are not equal to the expressions of energy density and the pressure defined using ordinary volume in LQC. It is straightforward to check that the  gauge-covariant energy density and pressure satisfy the conservation law 
\bq
\lb{energy conservation}
\dot \rho_{\mathrm{g.c.}}+3H_{\mathrm{g.c.}}\left(\rho_{\mathrm{g.c.}}+P_{\mathrm{g.c.}}\right)=0 ~.
\eq

On the contrary, if  the energy density and the pressure are defined with respect to the conventional volume via
\bq
\tilde \rho\equiv \frac{\mathcal H^{\mathrm{g.c.}}_m}{v}=\rho_0 v^{-w-1}\sinc^{-3w}\left(\frac{\lambda b}{2}\right), \quad \tilde P\equiv -\frac{\partial \mathcal H^\mathrm{\mathrm{g.c.}}_m}{\partial v}=w\tilde\rho,
\eq
then it can be shown that  due to the additional $\sinc$ terms, $\tilde \rho$ and $\tilde P$ do not satisfy the energy conservation law as 
\bq
\dot {\tilde \rho}+3 H\left(\tilde \rho+\tilde P\right)\neq  0,
\eq
with $H=\frac{\dot v}{3 v}$ denoting the conventional Hubble rate. Hence, from the perspective of the energy conservation law, it is natural to employ the gauge covariant quantities in  gLQC. 

In terms of the gauge covariant volume, the effective Hamiltonian constraint (\ref{gauge covariant ham}) takes the form
\bq
\lb{gauge covariant ham}
\mathcal H_{\mathrm{g.c.}} =-\frac{6v_\mathrm{g.c.}}{\kappa \beta^2\lambda^2}\sin^2\left(\lambda b\right)\sinc^{-2}\left(\frac{\lambda b}{2}\right) +\rho_0v^{-w}_\mathrm{g.c.} .
\eq
Note that even though the Hamiltonian constraint is expressed in terms of $v_\mathrm{g.c.}$, the phase space is still labelled by $b$ and $v$. The Hamilton's equations for phase space variables are:
\bqn
\lb{vdot-old}
\dot v &=&\frac{3v\sinc^2(\frac{\lambda b}{2})}{4b\beta}\Big\{b^2\cos(\frac{\lambda b}{2})\left(1+5\cos\left(\lambda b\right)-\frac{2}{b\lambda}\sin\left(b\lambda\right)\right)+\kappa \beta^2 P_{g.c.}\left(\cos(\frac{\lambda b}{2})-\frac{2}{b\lambda}\sin(\frac{\lambda b}{2})\right)\Big\},\nb\\
\lb{bdot}
\dot b &=&-4\pi G \beta \sinc^3\left(\frac{\lambda b}{2}\right)\left(\rho_{g.c.}+P_{g.c.}\right).
\eqn
These equations turns out to be far more complicated than those in LQC (see eg. \cite{Singh:2009mz, Agullo:2016tjh}). It turns out that due to $sinc$ term in the matter sector of the Hamiltonian constraint (\ref{gauge covariant ham}), the quantum geometry modified Friedmann and Raychaudhuri equations, in conventional volume, in  gLQC do not in general have closed forms.  Moreover, even using the gauge covariant volume $v_\mathrm{g.c.}$, it is still difficult to find a closed form of the Friedmann equation in terms of $v_\mathrm{g.c.}$. The reason is that the Hamiltonian constraint is still a transcendental equation which contains both $b$ and the trigonometric functions of $b$. The closed form is also difficult to obtain 
if one works with a phase space with gauge-covariant volume $v_\mathrm{g.c.}$ and its conjugate variable $b_\mathrm{g.c.}$,
\bq
b_\mathrm{g.c.}=\int \frac{\mathrm d b}{\sinc^3\left(\lambda b/2\right)},
\eq
with $\{b_\mathrm{g.c.},v_\mathrm{g.c.}\}=4\pi G\beta$. From the Hamilton's equations for $b$ and $v$ we find that the gauge-covariant volume and its conjugate variable satisfies the following dynamical equations 
\bqn
\lb{vdot}
\dot v_\mathrm{g.c.} &=&\frac{3v_\mathrm{g.c.}}{2\lambda \beta}\cos\left(\frac{\lambda b}{2}\right)\sinc^2\left(\frac{\lambda b}{2}\right)\Big[2\sin\left(\lambda b\right)+\lambda b\cos\left(\lambda b\right)-\lambda b\Big], \\
\dot b_\mathrm{g.c.} &=& -4\pi G \beta \left(\rho_{g.c.}+P_{g.c.}\right).
\eqn
The intricate nature of these equations and the fact that we can not replace $b$ with $b_\mathrm{g.c.}$ to make the Hamiltonian constraint non-transcendental, imply that 
 going to $(b_\mathrm{g.c.},v_\mathrm{g.c.})$ phase space offers no help to find the closed forms of quantum geometry modified Friedmann and Raychaudhuri equations. 
For this reason, it is important to investigate the asymptotic behavior of the Friedmann and Raychaudhuri equations in the classical regime and obtain the higher order corrections to the classical theory. In earlier works, it has been shown that GR is recovered in the large volume regime where spacetime curvature is extremely  small 
\cite{Liegener:2019ymd,Liegener:2019dzj}, however correction terms to the classical dynamics originating from quantum geometric effects were not understood. 
 
There is however, one exceptional case where one can find the modified Friedmann  equation in a closed form in gLQC. This special case deals with a positive cosmological constant which is discussed next.  

\subsection{The modified Friedmann equation in gLQC with a positive $\Lambda$}
As discussed above, a closed form of the modified Friedmann and Raychaudhuri equations is difficult to obtain in gLQC  due to the additional $sinc$ term in the Hamiltonian constraint arising from the gauge covariant fluxes. However, if the matter content consists only of a positive cosmological constant $(\Lambda)$,  one finds that a closed form of the Friedmann and Raychaudhuri equations becomes available. 

For a positive $\Lambda$, the equation of state $w = -1$ and the  energy density in the Hamiltonian constraint (\ref{gauge covariant ham}) is a constant which is given by 
\bq
\rho_0=\frac{2\Lambda}{\kappa}.
\eq
Then the vanishing of the Hamiltonian constraint leads to
\bq
\lb{constraint with lambda}
x^2 \cos^2\left(x\right)=\frac{\Lambda \beta^2 \lambda^2}{12},
\eq
with $x:=\lambda b/2$. Since the asymptotic regions for the classical limits correspond to those near $x=0$ in the expanding phase and $x=\pi/2$ in the contracting phase, one can restrict $x$ to the interval $x\in(0,\pi/2)$ and solve (\ref{constraint with lambda}) for any positive $\Lambda$. Denote the root of the constraint equation by $x_0$. Note the momentum $b$ is a constant of motion as can be seen from (\ref{bdot}) for $w=-1$. As a result, (\ref{constraint with lambda}) can be directly  solved, leading to 
\bq
\cos x_0=\frac{\beta \lambda }{x_0}\sqrt{\frac{\Lambda}{12}}, \quad \quad \sin x_0=\sqrt{1-\frac{\Lambda \beta^2 \lambda^2}{12x^2_0}}.
\eq
Substituting these equations into (\ref{vdot}), we get 
\bq
H^2_\mathrm{g.c.}=\frac{\tilde \Lambda}{3},
\eq
which has the same form as the classical Friedmann equation with an effective cosmological constant $\tilde \Lambda$. Note when $b$ is a constant, the conventional Hubble rate is equal to the gauge covariant Hubble rate. The rescaled $\tilde \Lambda$ is explicitly given by 
\bq
\tilde \Lambda=\frac{12}{\beta^2 \lambda^2 x^4_0}\sin^4(x_0)\cos^2(x_0)\Big[x_0-\sin\left(2x_0\right)\Big]^2.
\eq
The classical limit can be recovered when the cosmological constant $\Lambda$ is far below the Planck scale. Under this condition, $x_0\rightarrow \lambda \beta \sqrt{\Lambda/12}$ and thus $\tilde \Lambda \rightarrow \Lambda$.  The Raychaudhuri equation in the current case is simply given by 
\bq
\ddot a/a=H^2_\mathrm{g.c.}=\frac{\tilde \Lambda}{3} ~.
\eq
In contrast, when the momentum $b$ is time-dependent, the vanishing of  the Hamiltonian constraint (\ref{gauge covariant ham}) leads to  a transcendental equation which can not be solved analytically. As a result, a closed form of the Friedmann and Raychaudhuri equations can not be obtained. In this situation,  one can analytically analyze the asymptotic behavior of the Friedmann and Raychaudhuri equations in the classical limits in both pre-bounce (contracting) and post-bounce (expanding) branches as discussed in detail in the next section. 

\section{The weak curvature limit of the modified Friedmann and Raychaudhuri equations in  \MakeLowercase{g}LQC}
\label{sec:asymptotic gauge covariant LQC}
\renewcommand{\theequation}{4.\arabic{equation}}\setcounter{equation}{0}

In this section, we consider the asymptotic forms of the Friedmann and Raychaudhuri equations in the contracting (pre-bounce) and expanding (post-bounce) phases of gLQC in the large volume, weak curvature limit. It turns out that the quantum geometric corrections to classical   Friedmann and Raychaudhuri equations in the expanding and contracting phases do not coincide since two branches are asymmetric with respect to the bounce in gLQC. We first discuss the large volume expansion of the momentum $b$ based on the requirement for a vanishing Hamiltonian constraint  and then apply this expansion to the dynamical equations to derive the asymptotic forms of the Friedmann and Raychaudhuri equations in terms of the powers of the gauge-covariant energy density. Finally, in addition to the gauge-covariant Hubble rate, we also  give the  asymptotic form  of the Friedmann equation in terms of  the conventional Hubble rate and the energy density. 

\subsection{The asymptotic  Friedmann and Raychaudhuri  equations in the expanding phase}

In gLQC, the  gauge covariant energy density is related with the geometric degrees of freedom via  
\bq
\rho_\mathrm{g.c.}=\frac{6\sin^2\left(\lambda b\right)}{\kappa \beta^2\lambda^2\sinc^2\left(\lambda b/2\right)},
\eq
which implies that the classical regime can only be reached near $b=0$.  Although $\sin\left(\lambda b\right)$ is a periodic function with infinite zeroes, the presence of $\sinc$ term in the denominator in general rescales $\kappa$ at these turning points. The only regime in which $\rho_\mathrm{g.c.}\rightarrow 0$ without a rescaled $\kappa$ is $b\rightarrow 0$.  As  a result, $b\rightarrow 0$ corresponds to the classical limit in the expanding branch where $\kappa$ takes its observed value.  Moreover, from (\ref{bdot}), we know $b$ is a monotonically decreasing function in the forward evolution of time in a small regime $b>0$ close to $b=0$. As a result, the bounce would happen when $\rho_\mathrm{g.c}$ attains its maximum energy density at $b\approx 0.756798$, giving $\rho^\mathrm{max}_\mathrm{g.c.}\approx 0.515531$. The minimal positive $b$ for a vanishing $\rho_\mathrm{g.c.}$ happens at $b=\pi/\lambda$ which corresponds to the classical limit in the contracting branch before the bounce. When $b\rightarrow\pi/\lambda$, $\sinc^2\left(\lambda b/2\right) \rightarrow \frac{4}{\pi^2}$, so we expect a rescaled $\kappa$ (or equivalently a rescaled Newton's constant) in the distant past of the contracting branch. As a result, in the following, we  restrict $b$ to the interval $b\in\left(0, \pi/\lambda\right)$ and  discuss the asymptotic behavior of the Friedmann and Raychaudhuri equations at $b= 0^+$ and $b=\left(\pi/\lambda\right)^-$.

In the classical regimes when   $\rho_\mathrm{g.c.}\rightarrow 0$, the volume  $v_\mathrm{g.c.}$ is expected to be much larger than unity. As a result, one can expand the momentum $b$ in the powers of the inverse volume. The large volume expansion of $b$ in terms of $1/v_\mathrm{g.c.}$ is supposed to satisfy the Hamiltonian constraint at each consecutive order. More specifically, the vanishing of the Hamiltonian constraint (\ref{gauge covariant ham}) leads to the following expansion near $b=0$,
\bq
\lb{3a1}
-\frac{6b^2}{\kappa \beta^2}+\frac{3\lambda^2 b^4}{2\kappa \beta^2}-\frac{\lambda^4b^6}{8\kappa \beta^2}+\rho_0v^{-1-w}_\mathrm{g.c.}+\mathcal O\left(b^8\right)=0.
\eq
Noting that the lowest order in the above expansion is $b^2$ and there are only even powers of $b$, thus we assume an ansatz for the large volume expansion of $b$ in the expanding phase  near $b = 0$,
\bq
\lb{3a2}
b=v^{-\frac{1}{2}-\frac{w}{2}}_\mathrm{g.c.}\left(a_0+a_1 v^{-\alpha_1}_\mathrm{g.c.}+a_2 v^{-\alpha_2}_\mathrm{g.c.}+a_3 v^{-\alpha_3}_\mathrm{g.c.}+...\right),
\eq
where $a_0$, $a_i$ and $\alpha_i$ with $i=1,2,3...$ are parameters to be determined by the Taylor series (\ref{3a1}). We also require $\alpha_i$ to be positive and that their magnitudes increase with $i$ so that $b$ in (\ref{3a2}) is expanded in the ascending powers of the inverse volume.  
Now, plugging (\ref{3a2}) into (\ref{3a1}) and requiring that the constraint holds at each order of the expansion, we find that the first three powers of the inverse volume in the expansion turn out to be 
\bq
\lb{3a3}
\alpha_1=1+w,\quad \quad \alpha_2=2+2w,\quad \quad \alpha_3=3+3w .
\eq
Meanwhile, the expansion coefficients are
\bq
\lb{3a4}
a_0=\sqrt{\frac{\kappa \rho_0 \beta^2}{6}},\quad  a_1=\frac{\lambda^2}{8}a^3_0,\quad a_2=\frac{17\lambda^4}{384} a^5_0 .
\eq
The expression for $a_3$ is not shown explicitly since it is not required for computing $\rho^3_\mathrm{g.c.}$ terms in the  Friedmann and Raychaudhuri equations. Next we only need to expand the dynamical equation (\ref{vdot}) near $b\approx0$ and then use (\ref{3a2}), to obtain 
\bq
\lb{3a4}
\frac{\dot v_\mathrm{g.c.}}{3v_\mathrm{g.c.}}|_{b\approx0}=\sqrt{\frac{\kappa \rho_\mathrm{g.c.}}{6}}\left(1-\frac{\kappa \lambda^2\beta^2\rho_\mathrm{g.c.}}{12}-\frac{7\kappa^2 \beta^4\lambda^4\rho^2_\mathrm{g.c.}}{4320}\right)+\mathcal O\left(\rho^{7/2}_\mathrm{g.c.}\right) .
\eq
Here we have used $\rho_\mathrm{g.c.}=\rho_0/v^{1+w}_\mathrm{g.c.}$.  Correspondingly, the asymptotic form  of the square of the gauge covariant Hubble rate near $b \approx 0$ in the expanding phase reads 
\bqn
\lb{3a7}
H^2_\mathrm{g.c.}|_{b\approx0}&=&\frac{\kappa \rho_\mathrm{g.c.}}{6}-\frac{\kappa^2\beta^2\lambda^2\rho^2_\mathrm{g.c.}}{36}+\frac{\kappa^3\beta^4\lambda^4\rho^3_\mathrm{g.c.}}{1620}+\mathcal O\left(\rho^4_\mathrm{g.c.}\right),\nb\\
&=&\frac{\kappa \rho_\mathrm{g.c.}}{6}\left(1-\frac{\rho_\mathrm{g.c.}}{\rho^\mathrm{LQC}_\mathrm{max}}\right)+\mathcal O\left(\rho^3_\mathrm{g.c.}\right),
\eqn
where $\rho^\mathrm{LQC}_\mathrm{max}=\frac{3}{8\pi G\lambda^2\beta^2}$ is the maximum energy density in LQC.
Hence in gLQC, up to the second order in the gauge covariant energy density, the asymptotic form of the Friedmann equation in terms of the gauge covariant quantities takes exactly the same form as  the modified Friedmann equation in LQC with the same maximum energy density! As a result, since only the gauge-covariant quantities satisfy the conservation law and thus are regarded as the observables in gLQC, an observer living in such a universe will regard LQC as the first order correction beyond GR. 

It is important to note that in gLQC, the higher order corrections $\mathcal O\left(\rho^3_\mathrm{g.c.}\right)$ do not vanish identically. In contrast, one can also use the same techniques to compute the asymptotic expansion of the Friedmann equation in LQC in the large volume limit. Since the well-known Friedmann equation in LQC only contains higher-order terms up to $\rho^2$, one is bound to find that the resulting series expansion terminates at the next-to-leading order correction to GR, namely terms higher than $\rho^2$ all vanish identically. Finally, the nonvanishing of $\mathcal O\left(\rho^3_\mathrm{g.c.}\right)$ can also be inferred by noting that $\rho^\mathrm{LQC}_\mathrm{max}\approx 0.409374$ is less than the maximum covariant energy density $\rho^\mathrm{max}_\mathrm{g.c.}\approx 0.515531$ in gLQC. This implies that the higher order terms $\mathcal O\left(\rho^3_\mathrm{g.c.}\right)$ in (\ref{3a7}) also contribute to the maximum energy density in gLQC.  Therefore, recovering LQC dynamical equations in the low curvature limit of gLQC by no means implies that the bounce would take place at low curvature since higher-order terms will become important in the high curvature regime in which the first few terms in our series expansion can not capture the whole evolutionary dynamics.

In addition, it is also straightforward to compute the asymptotic form of the Friedmann equation in terms of the conventional Hubble rate and the energy density in gLQC by using (\ref{conversion between h}) and the relation 
\bq
\lb{conversion between rho}
\rho_\mathrm{g.c}=\frac{\rho}{\sinc^{3w+3}\left(\lambda b/2\right)} .
\eq
The result turns out to be 
\bq
\lb{3a5}
H^2|_{b\approx0}=\frac{\kappa \rho}{6}-\frac{9+w}{288}\kappa^2\beta^2\lambda^2\rho^2+\frac{1240-7w-30w^2}{414720}\kappa^3\beta^4\lambda^4\rho^3+\mathcal O\left(\rho^4\right).
\eq
Note in terms of the conventional energy density, the higher order terms in (\ref{3a5}) explicitly depends on the equation of state of the perfect fluid which is a consequence of the ``non-minimal" coupling between gravity and the matter sector in gLQC.

Apart from the Friedmann equation, one can also compute the asymptotic form of the Raychaudhuri equation in the expanding phase.  Two different approaches are available. First, one can use the equation of motion and note that $\ddot a_\mathrm{g.c.}= \{\dot a_\mathrm{g.c.}, \mathcal  H_\mathrm{g.c.}\}=\{\frac{\dot v_\mathrm{g.c.}}{3v^{2/3}_\mathrm{g.c.}},\mathcal H_\mathrm{g.c.}\}$. Since we already know $\dot v_\mathrm{g.c.}$ in (\ref{vdot}), one can further compute $\ddot a_\mathrm{g.c.}$ and then expand the resulting equation in the Taylor series near $b\approx 0$. Finally using (\ref{3a2}), one can truncate the Raychaudhuri equation at any desired order of $\rho_\mathrm{g.c.}$.  Alternatively, since the conservation law (\ref{energy conservation}) holds for all orders (non-perturbatively) one can use the Friedmann equation and the energy conservation law to  derive the Raychaudhuri equation up to any perturbative order. Combining the Friedmann equation (\ref{3a7}) in terms of the gauge covariant Hubble rate and the gauge covariant energy conservation law (\ref{energy conservation}), we can find the  asymptotic form of the Raychaudhuri equation in terms of the gauge covariant variables which turns out to be  
\bq
\lb{3a8}
\frac{\ddot a_\mathrm{g.c.}}{a_\mathrm{g.c.}}|_{b\approx 0}=-\frac{4\pi G}{3}\rho_\mathrm{g.c.}\left(1-4\frac{\rho_\mathrm{g.c.}}{\rho^\mathrm{LQC}_\mathrm{max}}\right)-4\pi G P_\mathrm{g.c.}\left(1-2\frac{\rho_\mathrm{g.c.}}{\rho^\mathrm{LQC}_\mathrm{max}}\right)+\mathcal O\left(\rho^3_\mathrm{g.c.}\right),
\eq
with $a_\mathrm{g.c.}=a \sinc\left(\lambda b/2\right)$.  As in the case of the Friedmann equation with quantum gravity corrections up to $\rho^2_\mathrm{g.c.}$,  (\ref{3a8}) coincides with the form of the modified Raychaudhuri equation in terms of $\rho$  and $P$ in LQC. The higher order terms $\mathcal O\left(\rho^3_\mathrm{g.c.}\right)$ in (\ref{3a8}) do not vanish while in LQC terms with powers higher than $\rho^2$ vanish identically.
 
\subsection{The asymptotic  Friedmann and Raychaudhuri  equations in the contracting phase}
The derivation of the asymptotic  forms of  the Friedmann and Raychaudhuri  equations in the pre-bounce (contracting phase)  proceeds in the same way as in the expanding phase.  The main difference is that in the contracting phase, we need to compute the Taylor expansion around $b\approx\pi/\lambda$ or equivalently $b_*\approx 0$, with $b=\frac{\pi}{\lambda}-b_*$. The Hamiltonian constraint (\ref{gauge covariant ham}) can thus be expanded in terms of powers of $b_*$ as 
\bq
\lb{3b1}
\frac{3}{\kappa \beta^2}\Bigg\{\frac{1}{2}\pi^2 b^2_*-\pi\lambda b^3_*+\frac{1}{2}\lambda^2 b^4_*-\frac{1}{24}\pi^2\lambda^2b^4_*\Bigg\}-\frac{\rho_0}{v^{1+w}}+\mathcal O\left(b^5_*\right)=0,
\eq
which in contrast to (\ref{3a1})  contains both even and odd powers of $b_*$. As a result, the large volume expansion of $b_*$ is
\bq
\lb{3b2}
b_*=v^{-\frac{1}{2}-\frac{w}{2}}_\mathrm{g.c.}\left(c_0+c_1 v^{-\frac{1}{2}-\frac{w}{2}}_\mathrm{g.c.}+c_2 v^{-1-w}_\mathrm{g.c.}+\mathcal O\left(v^{-\frac{3}{2}-\frac{3w}{2}}_\mathrm{g.c.}\right) \right).
\eq
Plugging the above expansion into (\ref{3b1}), the coefficients in (\ref{3b2}) turn out to be 
\bq
\lb{3b3}
c_0=\sqrt{\frac{2\kappa \rho_0\beta^2}{3\pi^2}},\quad c_1=\frac{\lambda}{\pi}c^2_0,\quad c_2=\left(\frac{1}{24}-\frac{1}{\pi^2}\right)\lambda^2c^3_0.
\eq
Then,  expanding the dynamical equation (\ref{vdot})  at $b_*=0$ and  plugging in (\ref{3b2}) and (\ref{3b3}),  we find 
\bq
\lb{3b4}
\frac{\dot v_\mathrm{g.c.}}{3v_\mathrm{g.c.}}|_{b_*\approx 0}=-\sqrt{\frac{\tilde \kappa \rho_\mathrm{g.c.}}{6}}-\frac{\tilde \kappa\beta \lambda \rho_\mathrm{g.c.}}{12}+\frac{\left(4+\pi^2\right)(\tilde \kappa \rho_\mathrm{g.c.})^{3/2}\beta^2\lambda^2}{48\sqrt{6}}+\mathcal O\left(\rho^2_\mathrm{g.c.}\right),
\eq
here $\tilde \kappa=\kappa\left(\frac{2}{\pi}\right)^4$ which implies a rescaled Newton's constant in the classical regime of the contracting phase in gLQC.  
The leading-order term in the above expansion is negative, which is consistent with the fact that the expansion is performed in the contracting phase where $\dot v_\mathrm{g.c.}<0$.
From (\ref{3b4}), it is straightforward to find the asymptotic form of  the square of the gauge covariant Hubble rate in the classical regime, which reads 
\bq
\lb{3b10}
H^2_\mathrm{g.c.}|_{b_*\approx 0}=\frac{\tilde \kappa }{6}\rho_\mathrm{g.c.}+\frac{\lambda \beta \tilde \kappa^{3/2}}{6\sqrt{6}}\rho^{3/2}_\mathrm{g.c.}-\frac{3+\pi^2}{144}\lambda^2\beta^2 \tilde \kappa^2\rho^2_\mathrm{g.c.}+\mathcal O\left(\rho^{5/2}_\mathrm{g.c.}\right).
\eq
Note that in terms of the gauge covariant Hubble rate and the gauge covariant energy density, the Newton's constant in the Friedmann equation (\ref{3b10}) is rescaled by a constant $16/\pi^4$,  which is independent of the equation of state of the perfect fluid. Besides, in (\ref{3b10}), in addition to the integer powers of the gauge covariant energy density, there are also half integer powers which are missing in the asymptotic forms of the Friedmann equation in the expanding phase. With the help of the energy conservation law (\ref{energy conservation}), one can  obtain  the Raychaudhuri  equation up to $\rho^2_\mathrm{g.c.}$ order
\bqn
\lb{3b6}
\frac{\ddot a_\mathrm{g.c.}}{a_\mathrm{g.c.}}|_{b_*\approx0}&=&-\frac{\tilde \kappa }{12}\rho_\mathrm{g.c.}\left(1+\frac{5\lambda \beta }{2\sqrt 6}\tilde \kappa^{1/2}\rho^{1/2}_\mathrm{g.c.}-\frac{3+\pi^2}{6}\lambda^2\beta^2\tilde \kappa \rho_\mathrm{g.c.}\right)\nb\\
&&-\frac{\tilde \kappa}{4}P_\mathrm{g.c.}\left(1+\frac{3\lambda \beta }{2\sqrt 6}\tilde \kappa^{1/2}\rho^{1/2}_\mathrm{g.c.}-\frac{3+\pi^2}{12}\lambda^2\beta^2\tilde \kappa\rho_\mathrm{g.c.}\right)+\mathcal O\left(\rho^{5/2}_\mathrm{g.c.}\right).
\eqn 
Similarly the modified Raychaudhuri  equation in the contracting phase of  gLQC  includes an additional $\rho^{3/2}_\mathrm{g.c.}$ term as compared with its counterpart in the expanding phase. The asymptotic form of the Friedmann and the Raychaudhuri  equations in two branches clearly shows an asymmetric bounce in gLQC. Besides, the rescaled Newton's constant does not depend on the equation of state of the matter if the Friedmann and the Raychaudhuri equations are expressed in terms of the gauge covariant variables. 

Finally, similar to the expanding phase, one can also obtain the asymptotic form of the Friedmann equation in terms of the conventional Hubble rate and the energy density in the contracting phase by using the relations (\ref{conversion between h}) and (\ref{conversion between rho}), the result is
\bq
\lb{3b5}
H^2|_{b_*\approx 0}=\frac{\bar \kappa}{6}\rho+\frac{\lambda \beta\bar \kappa^{3/2}}{6\sqrt{6}}\rho^{3/2}-\frac{\bar \kappa^2\beta^2\lambda^2 f_1(w)}{1152}\rho^2+\mathcal O\left(\rho^{5/2}\right),
\eq
where $\bar \kappa=\kappa\left(\frac{\pi}{2}\right)^{3w-1}$ and 
\bq
f_1(w)=54w^2+72w-6+11\pi^2+3w\pi^2.
\eq 
Note the $\rho^2$ term in (\ref{3b5}) is negative definite since $f_1$ is positive definite and attains its minimum value $f^\mathrm{min}_1\approx 54.77$ at $w\approx -0.94$.
The rescaled  $\kappa \rightarrow \bar \kappa$ implies a rescaled Newton's constant $\bar G= \left(\frac{\pi}{2}\right)^{3w-1} G$ at each perturbative order. The rescaling factor explicitly depends on the equation of state due to the non-minimal coupling between the matter and gravity in gLQC. Our result is consistent with the result for a massless scalar field in \cite{Liegener:2019ymd} where $w=1$ and thus  $\bar G= \frac{\pi^2 G}{4} $.  It should also be  noted that although there is a rescaled Newton's constant in the contracting phase, one can still reach the classical regime when $\rho\rightarrow 0$.  This is in contrast with another variant of LQC \cite{Li:2018opr} (so-called mLQC-I) where not only a rescaled Newton's constant emerges in the contracting phase but also an effective Planck-scale cosmological constant shows up as $\rho\rightarrow 0$ in the contracting phase. 

\section{Conclusions}
\label{sec:summary}
\renewcommand{\theequation}{4.\arabic{equation}}\setcounter{equation}{0}

A detailed study of the variants of standard LQC in loop cosmology, and their relationship, helps understand the robustness of the quantum geometry effects in the very early universe. Recently, some modified versions of LQC have been proposed and studied in detail for a better understanding of the robust features of the cosmological sector from full LQG \cite{Li:2021mop}.  In this manuscript, we have studied large volume, weak curvature limit of the dynamical equations in one of the variants of LQC, referred to as gLQC, and explored its relationship with LQC. Loop quantization in gLQC is based on the holonomies and the gauge covariant fluxes which are motivated by dealing with non-closure of Poisson bracket between triads in the full theory. This is in contrast to LQC based on holonomies and gauge-fixed triads. We have showed in this paper that due to a non-trivial form of modifications from gauge-covariant fluxes,  the closed-form expressions of the modified Friedmann and Raychaudhuri equations are generally not available in gLQC. The only exception is for the matter content which only consists of a positive cosmological constant. In this particular case, the resulting Friedmann equation takes its classical form with an effective cosmological constant which tends to its classical value in the weak curvature limit. In a general case, since the Hamiltonian constraint in gLQC amounts to a transcendental equation for the momentum $b$, one can not analytically solve it in terms of energy density, and as a result one can not obtain closed forms of the modified Friedmann and Raychaudhuri equations. 
Note that unavailability of these equations does not mean that there is an obstacle to extract dynamics, such as numerically using Hamilton's equations, but their availability certainly makes a comparison with LQC more transparent.
 
In spite of the  non-availability of the closed forms of the dynamical equations in gLQC, one can still analyze the quantum geometry corrections to GR in the low curvature regime where the volume of the universe becomes much larger than the Planck scale. Besides, since the energy conservation law only holds for the gauge covariant energy density and pressure, gauge covariant quantities turn out to be the only legitimate variables in gLQC. Therefore, we consider the weak curvature limit of the dynamical equations in gLQC in terms of the gauge covariant quantities when the matter sector is described by a perfect fluid with a constant equation of state (and satisfies weak energy condition). In the post-bounce branch of gLQC, the classical limit corresponds to $b \rightarrow 0$. We first derive the large volume expansion of $b$  near $b=0$ from the Hamiltonian constraint and then substitute it into the Hamilton's equation of the gauge covariant volume. In this way, we find an asymptotic expansion of the Friedmann equation in terms of the gauge covariant Hubble rate as a sum of integer powers of the gauge covariant energy density ($\rho_\mathrm{g.c.}$). The leading order term recovers the classical limit in GR while the next-to-leading order term is the $\rho^2_\mathrm{g.c.}$ term. Truncating at $\rho^2_\mathrm{g.c.}$ order, we immediately obtain the same form of the modified Friedmann equation as in  LQC with the same maximum energy density. In other words, LQC is recovered in gLQC as the first order correction beyond GR.  Correspondingly, due to the energy conservation law, the asymptotic expansion of the Raychaudhuri equation at the $\rho^2_\mathrm{g.c.}$ order also takes the same form as the one in LQC. Moreover, it is important to note that recovering LQC dynamical equations in the weak curvature limit does not imply the bounce would take place at low curvature in gLQC since the higher-order terms in the series expansion will become equally important when the energy density approaches the Planck scale.
 
Similarly, one can follow the same procedures to compute the asymptotic forms of the Friedmann and Raychaudhuri equations in the contracting phase by noting that the classical limit is reached in the neighborhood of $b=\pi/\lambda$. Correspondingly, one can obtain the large volume expansion of $b$ from the Hamiltonian constraint and then plug it into the Hamilton's equation of the gauge covariant volume. The resulting Friedmann equation contains both integer and half-integer powers of the gauge covariant energy density, in particular, when truncated to the  $\rho^2_\mathrm{g.c.}$ order, there is an additional  $\rho^{3/2}_\mathrm{g.c.}$ term  which signifies an asymmetric evolution with respect to the bounce in gLQC. Besides, there appears a rescaled Newton's constant in the classical regime of the contracting phase which is a constant when the Friedmann and Raychaudhuri equations are expressed in terms of the gauge covariant energy density. This rescaled Newton's constant  becomes explicitly dependent on the equation of state when the dynamical equations are expressed in terms of the conventional volume and the energy density.  
 
Compared with the complicated structure of the Friedmann and Raychaudhuri equations in gLQC (and also modified loop cosmologies, mLQC-I and mLQC-II \cite{Li:2018opr,Li:2018fco}), our studies show that LQC turns out to be one of the simplest loop quantizations of the cosmological spacetime as its dynamical equations only include $\rho^2$ terms. Incorporating more features from LQG by using different quantization prescriptions or gauge covariant fluxes takes us  beyond LQC by effectively adding corrections higher than $\rho^2$ terms.  While modified Friedmann dynamics of LQC can be recovered as the first order correction beyond GR in gLQC for the post-bounce branch,
a similar study to understand the relationship between LQC and  modified loop cosmologies  (mLQC-I and mLQC-II), which are again based on holonomies and triads,  does not reveal any such link. It is possible that the relationship we find between LQC and gLQC may be a more general feature emerging from various studies on recovering cosmological sector of LQG using gauge-covariant fluxes which tells us that  though LQC may be recovered at lower orders of truncation, it is important to study higher orders to have a more complete understanding of the underlying quantum geometry.

\section*{Acknowledgments}

This work is supported by the NSF grant PHY-2110207, and the National Natural Science Foundation of China (NNSFC) with grant 12005186.


\begin{thebibliography}{10}

\bibitem{Ashtekar:2011ni}
A.~Ashtekar and P.~Singh,
\newblock {\em Loop Quantum Cosmology: A Status Report},
\newblock Class. Quant. Grav. {\bf 28}, 213001 (2011), arXiv:1108.0893.

\bibitem{Singh:2009mz}
P.~Singh,
\newblock {\em Are loop quantum cosmos never singular?},
\newblock Class. Quant. Grav. {\bf 26}, 125005 (2009), arXiv:0901.2750.

\bibitem{Singh:2014fsy}
P.~Singh,
\newblock {\em Loop quantum cosmology and the fate of cosmological
  singularities},
\newblock Bull. Astron. Soc. India {\bf 42}, 121 (2014), arXiv:1509.09182.

\bibitem{Agullo:2016tjh}
I. Agullo and P. Singh, {\em Loop Quantum Cosmology}, in Loop Quantum Gravity: The First 30 Years, edited by A. Ashtekar and J. Pullin
(Wald Scientific, Singapore, 2017), arXiv:1612.01236.

\bibitem{Ashtekar:2006rx}
A.~Ashtekar, T.~Pawlowski, and P.~Singh,
\newblock {\em Quantum nature of the big bang},
\newblock Phys. Rev. Lett. {\bf 96}, 141301 (2006), arXiv:gr-qc/0602086.

\bibitem{Ashtekar:2006uz}
A.~Ashtekar, T.~Pawlowski, and P.~Singh,
\newblock {\em Quantum Nature of the Big Bang: An Analytical and Numerical
  Investigation. I.},
\newblock Phys. Rev. D {\bf 73}, 124038 (2006), arXiv:gr-qc/0604013.

\bibitem{Ashtekar:2006wn}
A.~Ashtekar, T.~Pawlowski, and P.~Singh,
\newblock {\em Quantum Nature of the Big Bang: Improved dynamics},
\newblock Phys. Rev. D {\bf 74}, 084003 (2006), arXiv:gr-qc/0607039.

\bibitem{Ashtekar:2007em}
A.~Ashtekar, A.~Corichi, and P.~Singh,
\newblock {\em Robustness of key features of loop quantum cosmology},
\newblock Phys. Rev. D {\bf 77}, 024046 (2008), arXiv:0710.3565.

\bibitem{Diener:2014mia}
P.~Diener, B.~Gupt, and P.~Singh,
\newblock {\em Numerical simulations of a loop quantum cosmos: robustness of
  the quantum bounce and the validity of effective dynamics},
\newblock Class. Quant. Grav. {\bf 31}, 105015 (2014), arXiv:1402.6613.

\bibitem{Diener:2017lde}
P.~Diener, A.~Joe, M.~Megevand, and P.~Singh,
\newblock {\em Numerical simulations of loop quantum Bianchi-I spacetimes},
\newblock Class. Quant. Grav. {\bf 34}, 094004 (2017), arXiv:1701.05824.

\bibitem{Singh:2018rwa}
P.~Singh,
\newblock {\em Glimpses of Space-Time Beyond the Singularities Using
  Supercomputers},
\newblock Comput. Sci. Eng. {\bf 20}, 26 (2018), arXiv:1809.01747.

\bibitem{Corichi:2011sd}
A.~Corichi and E.~Montoya,
\newblock {\em On the Semiclassical Limit of Loop Quantum Cosmology},
\newblock Int. J. Mod. Phys. D {\bf 21}, 1250076 (2012), arXiv:1105.2804.

\bibitem{Diener:2014hba}
P.~Diener, B.~Gupt, M.~Megevand, and P.~Singh,
\newblock {\em Numerical evolution of squeezed and non-Gaussian states in loop
  quantum cosmology},
\newblock Class. Quant. Grav. {\bf 31}, 165006 (2014), arXiv:1406.1486.

\bibitem{Ashtekar:2015iza}
A.~Ashtekar and B.~Gupt,
\newblock {\em Generalized effective description of loop quantum cosmology},
\newblock Phys. Rev. D {\bf 92}, 084060 (2015), arXiv:1509.08899.

\bibitem{Engle:2007qh}
J.~Engle,
\newblock {\em Relating loop quantum cosmology to loop quantum gravity:
  Symmetric sectors and embeddings},
\newblock Class. Quant. Grav. {\bf 24}, 5777 (2007), arXiv:gr-qc/0701132.

\bibitem{Brunnemann:2007du}
J.~Brunnemann and C.~Fleischhack,
\newblock {\em On the configuration spaces of homogeneous loop quantum
  cosmology and loop quantum gravity},
\newblock (2007), arXiv:0709.1621.

\bibitem{Alesci:2013xd}
E.~Alesci and F.~Cianfrani,
\newblock {\em Quantum-Reduced Loop Gravity: Cosmology},
\newblock Phys. Rev. D {\bf 87}, 083521 (2013), arXiv:1301.2245.

\bibitem{Dapor:2017rwv}
A.~Dapor and K.~Liegener,
\newblock {\em Cosmological Effective Hamiltonian from full Loop Quantum
  Gravity Dynamics},
\newblock Phys. Lett. B {\bf 785}, 506 (2018), arXiv:1706.09833.

\bibitem{Han:2019feb}
M.~Han and H.~Liu,
\newblock {\em Improved $\overline{\mu}$-scheme effective dynamics of full loop
  quantum gravity},
\newblock Phys. Rev. D {\bf 102}, 064061 (2020), arXiv:1912.08668.

\bibitem{Yang:2009fp}
J.~Yang, Y.~Ding, and Y.~Ma,
\newblock {\em Alternative quantization of the Hamiltonian in loop quantum
  cosmology II: Including the Lorentz term},
\newblock Phys. Lett. B {\bf 682}, 1 (2009), arXiv:0904.4379.

\bibitem{Li:2018opr}
B.-F. Li, P.~Singh, and A.~Wang,
\newblock {\em Towards Cosmological Dynamics from Loop Quantum Gravity},
\newblock Phys. Rev. D {\bf 97}, 084029 (2018), arXiv:1801.07313.

\bibitem{Assanioussi:2018hee}
M.~Assanioussi, A.~Dapor, K.~Liegener, and T.~Paw\l{}owski,
\newblock {\em Emergent de Sitter Epoch of the Quantum Cosmos from Loop Quantum
  Cosmology},
\newblock Phys. Rev. Lett. {\bf 121}, 081303 (2018), arXiv:1801.00768.

\bibitem{Liegener:2019dzj}
K.~Liegener and P.~Singh,
\newblock {\em Some physical implications of regularization ambiguities in
  SU(2) gauge-invariant loop quantum cosmology},
\newblock Phys. Rev. D {\bf 100}, 124049 (2019), arXiv:1908.07543.

\bibitem{Liegener:2019ymd}
K.~Liegener and P.~Singh,
\newblock {\em New Loop Quantum Cosmology Modifications from Gauge-covariant
  Fluxes},
\newblock Phys. Rev. D {\bf 100}, 124048 (2019), arXiv:1908.07001.

\bibitem{Liegener:2019zgw}
K.~Liegener and P.~Singh,
\newblock {\em Gauge-invariant bounce from loop quantum gravity},
\newblock Class. Quant. Grav. {\bf 37}, 085015 (2020), arXiv:1906.02759.

\bibitem{Bianchi:2010zs}
E.~Bianchi, C.~Rovelli, and F.~Vidotto,
\newblock {\em Towards Spinfoam Cosmology},
\newblock Phys. Rev. D {\bf 82}, 084035 (2010), arXiv:1003.3483.

\bibitem{Gielen:2013naa}
S.~Gielen, D.~Oriti, and L.~Sindoni,
\newblock {\em Homogeneous cosmologies as group field theory condensates},
\newblock JHEP {\bf 06}, 013 (2014), arXiv:1311.1238.

\bibitem{Agullo:2018wbf}
I.~Agullo,
\newblock {\em Primordial power spectrum from the Dapor-Liegener model of loop
  quantum cosmology},
\newblock Gen. Rel. Grav. {\bf 50}, 91 (2018), arXiv:1805.11356.

\bibitem{Li:2018fco}
B.-F. Li, P.~Singh, and A.~Wang,
\newblock {\em Qualitative dynamics and inflationary attractors in loop
  cosmology},
\newblock Phys. Rev. D {\bf 98}, 066016 (2018), arXiv:1807.05236.

\bibitem{Li:2019ipm}
B.-F. Li, P.~Singh, and A.~Wang,
\newblock {\em Genericness of pre-inflationary dynamics and probability of the
  desired slow-roll inflation in modified loop quantum cosmologies},
\newblock Phys. Rev. D {\bf 100}, 063513 (2019), arXiv:1906.01001.

\bibitem{Li:2019qzr}
B.-F. Li, P.~Singh, and A.~Wang,
\newblock {\em Primordial power spectrum from the dressed metric approach in
  loop cosmologies},
\newblock Phys. Rev. D {\bf 101}, 086004 (2020), arXiv:1912.08225.

\bibitem{Li:2020mfi}
B.-F. Li, J.~Olmedo, P.~Singh, and A.~Wang,
\newblock {\em Primordial scalar power spectrum from the hybrid approach in
  loop cosmologies},
\newblock Phys. Rev. D {\bf 102}, 126025 (2020), arXiv:2008.09135.

\bibitem{Li:2021mop}
B.-F. Li, P.~Singh, and A.~Wang,
\newblock {\em Phenomenological implications of modified loop cosmologies: an
  overview},
\newblock Front. Astron. Space Sci. {\bf 8}, 701417 (2021), arXiv:2105.14067.

\bibitem{Li:2021fmu}
B.-F. Li and P.~Singh,
\newblock {\em Loop quantum gravity effects might restrict a cyclic evolution},
\newblock Phys. Rev. D {\bf 105}, 046013 (2022), arXiv:2108.12553.

\bibitem{Ashtekar:1998ak}
A.~Ashtekar, A.~Corichi, and J.~A. Zapata,
\newblock {\em Quantum theory of geometry III: Noncommutativity of Riemannian
  structures},
\newblock Class. Quant. Grav. {\bf 15}, 2955 (1998), arXiv:gr-qc/9806041.

\bibitem{Thiemann:2000bv}
T.~Thiemann,
\newblock {\em Quantum spin dynamics (QSD): 7. Symplectic structures and
  continuum lattice formulations of gauge field theories},
\newblock Class. Quant. Grav. {\bf 18}, 3293 (2001), arXiv:hep-th/0005232.

\end{thebibliography}
\end{document}